\documentclass{PoS}

\title{Spectrum of quarks in QCD$_{\mathbf{2}}$}

\ShortTitle{Spectrum of quarks in QCD$_2$}

\author{\speaker{H. Sazdjian}\\
        Institut de Physique Nucl\'eaire, CNRS/IN2P3\\
        Universit\'e Paris-Sud 11\\
        F-91405 Orsay, France\\
        E-mail: \email{sazdjian@ipno.in2p3.fr}}

\newcommand{\bc}{\begin{center}}
\newcommand{\ec}{\end{center}}
\newcommand{\be}{\begin{equation}}
\newcommand{\ee}{\end{equation}}
\newcommand{\bea}{\begin{eqnarray}}
\newcommand{\eea}{\end{eqnarray}}
\newcommand{\ba}{\begin{array}}
\newcommand{\ea}{\end{array}}
\newcommand{\lb}{\label}

\newcommand{\bfg}{\begin{figure}[htbp]}
\newcommand{\efg}{\end{figure}}
\newcommand{\pr}{Phys. Rev. }
\newcommand{\np}{Nucl. Phys. }

\newcommand{\prp}{Phys. Rep. }

\newcommand{\pl}{Phys. Lett. }

\abstract{The properties of the gauge invariant two-point quark 
Green's function are studied in the large-$N_c$ limit of 
two-dimensional QCD. The analysis is done by means of an exact 
integro\-differential equation. The Green's function is found to 
be infrared finite, with singularities in the momentum squared 
variable represented by an infinite number of threshold type 
branch points with a power -3/2, starting at positive mass squared 
values, with cuts lying on the positive real axis. The
expression of the Green's function is analytically determined.}

\FullConference{The 2011 Europhysics Conference on High Energy 
Physics-HEP 2011,\\	
July 21-27, 2011\\
Grenoble, Rh\^one-Alpes France}

\begin{document}

We report in this talk results concerning the gauge invariant 
two-point quark Green's function \cite{m,nm} considered in the 
large $N_c$ limit of two-dimensional QCD \cite{thft1,thft2,ccg}.
\par 
Defining the gauge invariant quark Green's function with a 
path-ordered gluon field phase factor along a straight line segment,
one can establish an exact integro\-differential equation for the
latter (in general in four-dimensional QCD and for any $N_c$),
where the kernel is made of a series of Wilson loops \cite{w,mgd,mk} 
along polygonal contours with a certain number of functional
derivatives acting on the contours \cite{s1}.
\par
Many simplifications occur in two-dimensional QCD at large $N_c$.
This theory is expected to have the essential features of confinement 
observed in four dimensions, with the additional simplification that 
asymptotic freedom is realized here in a trivial way, since the theory 
is super\-renormalizable. For simple contours, Wilson loop averages
in two-dimensional Yang-Mills theory are exponential functionals of 
the areas enclosed by the contours \cite{kzkk,kzk,br}. Furthermore, 
at large $N_c$, crossed diagrams and quark loop contributions disappear.
\par
It turns out that in two dimensions and at large $N_c$ the action
of the kernel of the above quoted integro\-differential equation
can explicitly be evaluated. The equation then reduces to the 
following form \cite{s2}:
\bea \lb{e1}
& &(i\gamma.\partial-m)S(x)=i\delta^2(x)
-\sigma\gamma^{\mu}(g_{\mu\alpha}g_{\nu\beta}-g_{\mu\beta}g_{\nu\alpha}) 
x^{\nu}x^{\beta}\nonumber \\
& &\ \ \ \ \ \times\left[\,\int_0^1d\lambda\,\lambda^2\,S((1-\lambda)x)
\gamma^{\alpha}S(\lambda x)
+\int_1^{\infty}d\xi\,S((1-\xi)x)\gamma^{\alpha}S(\xi x)\,\right], 
\eea
where $S$ is the gauge invariant quark Green's function with a
phase factor along a straight line segment, $x$ the ralative 
coordinate of the quark and antiquark fields and $\sigma$ the string 
tension.
\par
The equation is solved by decomposing $S$ into Lorentz invariant
parts, here in momentum space: 
\be \lb{e2}
S(p)=\gamma.pF_1(p^2)+F_0(p^2),
\ee
or, in $x$-space:
\be \lb{e3}
S(x)=\frac{1}{2\pi}\Big(\frac{i\gamma.x}{r}\widetilde F_1(r)
+\widetilde F_0(r)\Big),\ \ \ \ \ r=\sqrt{-x^2}. 
\ee
\par
One obtains, with the introduction of the Lorentz invariant functions,
two coupled equations. Their resolution proceeds through several steps, 
mainly based on the analyticity properties resulting from the
spectral representation of $S$ \cite{s1,s2}. The solutions are obtained 
in explicit form for any value of the quark mass $m$.
\par
The covariant functions $F_1(p^2)$ and $F_0(p^2)$ are, for complex $p^2$:
\bea
\lb{e4}
& &F_1(p^2)=-i\frac{\pi}{2\sigma}\,\sum_{n=1}^{\infty}\,
b_n\,\frac{1}{(M_n^2-p^2)^{3/2}},\\
\lb{e13}
& &F_0(p^2)=i\frac{\pi}{2\sigma}\,\sum_{n=1}^{\infty}\,
(-1)^nb_n\,\frac{M_n}{(M_n^2-p^2)^{3/2}}.
\eea
The masses $M_n$ ($n=1,2,\ldots$) have positive values greater than
the quark mass $m$ and are labelled with increasing values with respect 
to $n$; their squares represent the locations of branch point 
singularities with power $-3/2$, with cuts lying on the positive real 
axis of the complex plane of $p^2$. The masses $M_n$ and the coefficients
$b_n$ satisfy an infinite set of coupled algebraic equations that are
solved numerically. Their asymptotic behaviors for large $n$, such that
$\sigma\pi n\gg m^2$, are:
\be \lb{e5}
M_n^2\simeq \sigma\pi n,\ \ \ \ \ \ \ 
b_n\simeq \frac{\sigma^2}{M_n}.
\ee
\par
In $x$-space, the solutions are:
\be \lb{e15}
\widetilde F_1(r)=\frac{\pi}{2\sigma}\,\sum_{n=1}^{\infty}\,
b_n\,e^{-M_nr},\ \ \ \ \
\widetilde F_0(r)=\frac{\pi}{2\sigma}\,\sum_{n=1}^{\infty}\,
(-1)^{n+1}b_n\,e^{-M_nr}.
\ee
\par
At high energies, the solutions satisfy asymptotic freedom \cite{pltz}.
\par
In conclusion, the spectral functions of the quark Green's function 
are infrared finite and lie on the positive real axis of $p^2$. No 
singularities in the complex plane or on the negative real axis have 
been found. This means that the quarks contribute like physical particles 
with positive energies. (In two dimensions there are no physical gluons.)
\par 
The singularities of the Green's function are represented by an
infinite number of threshold type singularities, characterized by 
a power of $-3/2$ and positive masses $M_n$ ($n=1,2,\ldots$). The
corresponding singularities are stronger than simple poles and this 
feature might be at the origin of the unobservability of quarks as
asymptotic states. 
\par
The threshold masses $M_n$ represent dynamically generated masses,
since they are not present in the QCD Lagrangian. They survive 
even when the quark mass is zero. They play the role of gauge
invariant effective masses of quarks.
\par
\vspace{0.25 cm}

\noindent
\textbf{Acknowledgements.}
This work was supported in part by the European Community 
Research Infrastructure Integrating Activity ``Study of Strongly 
Interacting Matter'' (acronym HadronPhysics2, Grant Agreement 
No. 227431), under the Seventh Framework Programme of EU.
\par

\end{document}